# The 2019/20 Australian wildfires generated a persistent smoke-charged vortex rising up to 35 km altitude

*Sergey Khaykin[1], Bernard Legras[2], Silvia Bucci[2], Pasquale Sellitto[3], Lars Isaksen[4], Florent Tencé[1], Slimane Bekki[1], Adam Bourassa[5], Landon Rieger[5], Daniel Zawada[5], Julien Jumelet[1], Sophie Godin-Beekmann[1]*

[1] Laboratoire Atmosphères, Milieux, Observations Spatiales, UMR CNRS 8190, IPSL, Sorbonne Univ./UVSQ, Guyancourt, France.
[2] Laboratoire de Météorologie Dynamique, UMR CNRS 8539, IPSL, PSL-ENS/Sorbonne Univ./Ecole Polytechnique, Paris, France.
[3] Laboratoire Interuniversitaire des Systèmes Atmosphériques, UMR CNRS 7583, IPSL, Université Paris-Est Créteil/Université de Paris, Créteil, France.
[4] European Centre for Medium-Range Weather Forecasts, Reading, UK
[5] Institute of Space and Atmospheric Studies, University of Saskatchewan, Canada

**Abstract**

The Australian bushfires around the turn of the year 2020 generated an unprecedented perturbation of stratospheric composition, dynamical circulation and radiative balance. Here we show from satellite observations that the resulting planetary-scale blocking of solar radiation by the smoke is larger than any previously documented wildfires and of the same order as the radiative forcing produced by moderate volcanic eruptions. A striking effect of the solar heating of an intense smoke patch was the generation of a self-maintained anticyclonic vortex measuring 1000 km in diameter and featuring its own ozone hole. The highly stable vortex persisted in the stratosphere for over 13 weeks, travelled 66,000 km and lifted a confined bubble of smoke and moisture to 35 km altitude. Its evolution was tracked by several satellite-based sensors and was successfully resolved by the European Centre for Medium-Range Weather Forecasts operational system, primarily based on satellite data. Because wildfires are expected to increase in frequency and strength in a changing climate, we suggest that extraordinary events of this type may contribute significantly to the global stratospheric composition in the coming decades.

# Introduction

The impact of wildfire-driven thunderstorms on the global stratosphere has been deemed small until the North American wildfires in August 2017. Pyro-cumulonimbus (pyroCb) clouds from that event caused stratospheric perturbations an order of magnitude larger than the previous benchmarks of extreme pyroCb activity and approached the effect of moderate volcanic eruption[1,2]. Volcanic eruptions inject ash and sulphur which is oxidized and condenses to form submicron-sized aerosol droplets in the stratosphere. With the PyroCb, intense fire-driven convection lifts combustion products in gaseous form as well as particulate matter including organic and black carbon, smoke aerosols and condensed water. The solar heating of the highly absorptive black carbon propels the smoke-laden air parcels upward[1], which, combined with horizontal transport[3,4], leads to a more efficient meridional dispersion of these aerosols and prolongs their stratospheric residence time[5].

The Australian bushfires that raged in December 2019 – January 2020 have put a new benchmark on the magnitude of stratospheric perturbations. In this study we use various satellite observations to quantify the magnitude of hemispheric-scale perturbation of stratospheric gaseous compounds and aerosol loading caused by these wildfires. The radiative forcing of the stratospheric smoke is estimated using a radiative transfer model supplied by satellite observations of aerosol optical properties. Finally, using the operational forecasting system of the European Centre for Medium-Range Weather Forecasts (ECMWF)[6] we show that the solar heating of an intense smoke patch has led to generation of a quasi-ellipsoidal anticyclonic vortex which lofted a confined bubble of carbonaceous aerosols and water vapour up to 35 km altitude in about three months.

## Results

**Large-scale perturbation of the stratosphere**

The Australian wildfire season 2019/2020 was marked by an unprecedented burn area of 5.8 million hectares (21% of Australia's temperate forests)[7] and exceptionally strong PyroCb activity in the south-east of the continent[8]. The strongest PyroCb outbreak occurred on the New Year's Eve (Fig. 1a) and on the 1st of January the instantaneous horizontal extent of the stratospheric cloud amounted to 2.5 million km$^2$ as inferred from nadir-viewing TROPOMI[9] satellite measurement (Fig. 1b). On that day, an opaque cloud of smoke was detected in the stratosphere by the CALIOP space-based laser radar (lidar)[10] at altitudes reaching 17.6 km (Fig. 2). Another PyroCb outbreak with stratospheric impact, although less vigorous, took place on 4 January 2020 and on 7 January, the horizontal extent of the stratospheric smoke cloud peaked at 6.1 million km$^2$ (Fig. 1b) extending over much of the Southern midlatitudes (Fig. 2 and Supplementary Fig. 4c).

The high-altitude injections of smoke rapidly tripled the stratospheric aerosol optical depth (SAOD) in the southern extra tropics. The SAOD perturbation has by far exceeded the effect on stratospheric aerosol load produced by the North American wildfires in 2017, putting the Australian event on par with the strongest volcanic eruptions in the last 25 years (Fig. 3), i.e. since the leveling off of stratospheric aerosol load after a major eruption of Mount Pinatubo in 1991[12]. Three months after the PyroCb event, the SAOD perturbation has remained at the volcanic levels, gradually decreasing with a rate similar to the decay of stratospheric aerosol produced by moderate volcanic eruptions.

Using aerosol extinction profiles retrieved from the limb-viewing NASA OMPS-LP instrument[13] we find the total aerosol particle mass lofted into the so-called stratospheric "overworld"[14] (above 380 K isentropic level corresponding to ~12-17 km altitude) is 0.4±0.2 Tg (Fig. 4), which is nearly three times larger than the estimates for the previous record-high North American wildfires[2]. The increase in the stratospheric abundance of the gaseous combustion products, derived from the NASA Microwave Limb Sounder (MLS) satellite observations[16], is as remarkable as the aerosol increase. Fig. 4 puts in evidence that the stratospheric masses of carbon monoxide (CO) and acetonitrile ($CH_3CN$) bounded within the southern extra-tropics increase abruptly by 1.5±0.9 Tg (~20% of the pre-event levels) and 3.7±2.0 Gg (~5%), respectively, during the first week of 2020. The injected mass of water was estimated at 27±10 Tg that is about 3% of the total mass of stratospheric overworld water vapour in the southern extratropics (see Methods).

The gases and particles injected by the PyroCbs were advected by the prevailing westerly winds in the lower stratosphere. The patches of smoke dispersed across all of the Southern hemisphere extra-tropics in less than two weeks with the fastest patches returning back over Australia by 13 January 2020, whereas the carbon-rich core remained bounded within midlatitudes as shown in Fig. 2. During the following months, most of the particulate material dwelled in the lower stratosphere, the larger and heavier particles sedimented to lower altitudes while the carbon-rich fraction ascended from 15 to 35 km due to solar heating of black carbon (Supplementary Figure 1).

**Radiative forcing**

The large amount of aerosols produced a significant radiative forcing (RF), which we quantified using explicit radiative transfer modelling based on the measured aerosol optical properties (Supplementary notes 1). In the latitude band between 25-60°S, an average cloud-free reference monthly radiative forcing as large as about -1.0 W/m² at the top of the atmosphere (TOA) and -3.0 W/m² at the surface is found in February 2020 (Supplementary Figure 2). This can be attributed to perturbation to the stratospheric aerosol layer by the Australian fires plumes. The area-weighted global-equivalent cloud-free RF is estimated (Supplementary table 1) to values as large as -0.31±0.09 W/m$^2$ (TOA) and -0.98±0.17 W/m$^2$ (at the surface). It is important to notice that these estimations don't take the presence of clouds into account and are to be taken as purely reference values. For typical average cloud cover in the area affected by the plume[17], the surface all-sky RF can be reduced to ~50% and the TOA all-sky RF to ~30-50% of the clear-sky RF estimations[18] (see Supplementary notes 1 for details). From the perspective of the stratospheric aerosol layer perturbation, the global TOA RF produced by the Australian fires 2019/2020 is larger than the RF produced by all documented wildfire events and of the same order of magnitude of moderate volcanic eruptions during the last three decades (that have an integrated effect estimated at[19] -0.19±0.09 W/m$^2$, or smaller[20]). In contrast to the non-absorbing volcanic sulphates, the carbonaceous wildfire aerosols absorb the incoming solar radiation, leading to yet more substantial radiative forcing at the surface, due to the additional large amount of energy absorbed in the plume. This can be linked to the ascent of a smoke cloud in the stratosphere. This is discussed in the next section.

**Rising bubble of smoke**

The primary patch of smoke originating from the New Year's Eve PyroCb event followed an extraordinary dynamical evolution. By the 4 January 2020, en route across the Southern Pacific, the core plume started to encapsulate into a compact bubble-like structure, which was identified using CALIOP observations on 7 January 2020 as an isolated 4-km tall and 1000 km wide structure (Fig. 5a). Over the next 3 months, this smoke bubble crossed the Pacific and hovered above the tip of South America for a week. It then followed a 10-week westbound round-the-world journey that could be tracked until the beginning of April 2020 (Supplementary notes 2, Supplementary Figure 3), travelling over 66,000 km.

The large amount of sunlight-absorbing black carbon contained in the smoke cloud provided a localized heating that forced the air mass to rise through the stratosphere. With an initial ascent rate of about 0.45 km/day, the bubble of aerosol continuously ascended during the three months with an average rate of 0.2 km/day. While remaining compact, the bubble was leaking material from its bottom part, leaving an aerosol trail that was progressively dispersed and diluted, filling the whole mid-latitude austral stratosphere up to 30 km (Supplementary Figure 1, see also Fig. 10b). The rise ceased in late March 2020 when the top of the bubble reached 36 km altitude (Fig. 5b, Supplementary Figure 3). This is substantially higher than any coherent volcanic aerosol or smoke plume observed since the major eruption of Pinatubo in 1991.

Along with the carbonaceous aerosols, the bubble entrained tropospheric moisture in the form of ice aggregates injected by the overshooting PyroCbs. In the warmer stratosphere, the ice (detected by MLS sensor as high as 22 km, cf. Fig. 5b) eventually evaporated, enriching the air mass with water vapour. This led to extraordinary high water vapour mixing ratios emerging across the stratosphere within the rising smoke bubble (Fig.5b). The decay of CO within the bubble (Fig. 5c) was faster than that of water vapour, reflecting the fact that, unlike water vapour, the carbon monoxide was subject to photochemical oxidation whose efficiency increases sharply with altitude[21].

Temperature profiles from GNSS radio occultation sensors exhibit a clear dipolar anomaly within the bubble with a warm pole at its bottom and a cold pole at its top (Fig. 5d). Although counterintuitive from the pure radiative transfer perspective, the observed temperature dipole within the heated cloud represents an expected thermal signature of a synoptic-scale vortex.

**The vortex**

The compact shape of the smoke bubble could only be maintained through an efficient confinement process. The meteorological analysis of the real-time operational ECMWF integrated forecasting system (IFS)[6] reveals that a localized anticyclonic vortex was associated with the smoke bubble during all its travel, moving and rising with it (Figs. 5, 6a-b & Supplementary Figure 5). With a peak vorticity of $10^{-4}$ s$^{-1}$ (Figs. 6c & 7b) and a maximum anomalous wind speed of 13 m s$^{-1}$ (Fig. 6d) during most of its lifetime, the vortex had a turnover time of about 36 hours. It has therefore survived about 60 turnover times demonstrating a remarkable stability and resilience against perturbations. The ascent was surprisingly linear in potential temperature at a rate of 5.94±0.07 K day$^{-1}$ (Fig. 7a). This corresponds to a heating rate $dT/dt$ which varies from about 3 K day$^{-1}$ at the beginning of January to 1.5 K day$^{-1}$ at the end of March. The altitude rise is from 16 to 33 km for the vortex centroid and from 17 to 36 km for the top of the bubble according to CALIOP. The upper envelope of the OMPS detection of the bubble is seen as the cyan curve on Fig. 7a. This envelope is always above the top detected from CALIOP. Such a bias is expected as OMPS-LP is a limb instrument that scans a much wider area than the narrow CALIOP track. Both CALIOP and OMPS-LP detect that the top of the bubble rises initially faster than the vortex core, by about 10 K day$^{-1}$. This period corresponds to the initial travel of the bubble to the tip of South America. In terms of altitude ascent, the rates 10 and 5.94 K day$^{-1}$ translate approximately as 0.45 and 0.2 km day$^{-1}$.

The confining properties of the vortex are confirmed by the co-located anomalies in tracers and aerosol from the TROPOMI instrument. The satellite observations (Supplementary notes 3.3 & Supplementary Figure 4a,b) reveals an isolated enhancement of the aerosol absorbing index and of the CO columnar content, as well as the presence of a deep mini ozone hole, depleted by up to 100 DU. All three features were captured in the same position by the ECMWF analyses (Fig. 6e & Supplementary Figures 4 and 7).

**Companion vortices**

It is worth noticing that the vortex was not a single event. It had several companions, albeit of smaller magnitude and duration, also caused by localized smoke clouds. The most noticeable lasted one month and travelled the hemisphere. Another one found a path across Antarctica where it was subject to the strong aerosol heating of permanent daylight and rose up to 27 km.

The second vortex is borne from the smoke cloud that found its way to the stratosphere during the PyroCb event of 4-5 January 2020. This cloud initially travelled north east passing north of New Zealand before taking a south easterly direction crossing the path of the cloud emitted on 31 December 2019 and the main vortex. Fig. 8 a-b shows that a vortex-like structure can be spotted as early as 7 January, coinciding with the location of a compact bubble according to CALIOP. Subsequently the bubble crossed the path of the first vortex on 16 January while rising and intensifying (see Fig. 8 c-d) and travelled straight eastward crossing the Atlantic and the Indian Ocean until it reached the longitude of Australia by the end of January where it disappeared after travelling all the way round the globe. During this travel, the altitude of the vortex centroid rose from 15 to 19 km and the top of the bubble, as seen from CALIOP, reaches up to 20 km.

The third bubble has been first detected by CALIOP on 7 January at 69° S / 160° W. It then moved over Antarctica (Fig. 9a) until the end of January where it spent a week over the Antarctic Peninsula before moving to the tip of South America, shortly after this region was visited by the main vortex. It eventually moved to the Atlantic where it dissipated by 25 February (Fig. 9b). The bubble was accompanied by a vortex during its whole life cycle as seen in Fig. 9c. Although the magnitude of this vortex was modest compared to the main one and even the second one in terms of maximum vorticity (Fig. 9e), it performed a very significant ascent from 18 to 26 km (Fig. 9d). We attribute this effect to the very effective aerosol heating received during the essentially permanent daylight of the first period over Antarctica. The simultaneous rise of the main vortex and the third vortex is very clear from the OMPS-LP latitude-altitude sections in Fig. 10c-d.

The combined trajectories of the main vortex and its two companions are shown in Fig. 10e. Fig. 10a shows how the trajectories of the vortices form the skeleton of the dispersion path of the smoke plume in the stratosphere.

Fig. 10b shows that OMPS-LP follows closely the evolution of the top altitude of all the vortices under the shape of well-defined branches in a longitude-time Hovmöller diagram. It is worth noticing that both the main and the third vortex spent some time wandering in the vicinity of the Drake passage at the beginning of February. Such a stagnation situation is prone to sensitivity. The IFS forecasts during the end of January predicted that the main vortex would cross to the Atlantic, while instead it did not and began moving westward over the Pacific as it reached a higher altitude where easterly winds prevail. Several secondary branches that seem to separate from the main one followed by the main vortex are also visible in this diagram. A detailed inspection reveals that they are indeed associated with patches left behind by the main bubble as it moved upward. It is apparent from several of the panels of Fig. 7a that the top part of the bubble remained always compact while the bottom part was constantly leaking material. Fig. 10c-d shows latitude altitude cross sections of OMPS aerosols on 16 January and 1 February. The fast rise of the main vortex during that period is visible near 50S while the second vortex is located at lower altitude and the third vortex corresponds to the towering structure by 75-80S.

## Discussion

Long-lived anticyclones have also been observed as very rare events in the summer Arctic stratosphere[22, 23, 24], but they are much larger structures of about 2000 km, very near the pole, and do not display any of the specific characters of the self-generated smoke-charged vortex. According to geophysical fluid dynamics theory[25], a local heating in the austral stratosphere is expected to produce positive potential vorticity aloft and destroy it beneath. The positive potential vorticity is partially realized as anticyclonic rotation and partially as temperature stratification. The negative potential vorticity is apparently dispersed away with the tail of the bubble. The ECMWF analysed thermal structure (Figs. 6f & Supplementary Figure 7c) shows the same dipole as in the GNSS-RO satellite profiles with the same amplitude. The observed vortex is quite similar to known isolated ellipsoidal solutions of the quasi-geostrophic equations[26] which explain some of the long-lived vortices in the ocean[27], but such a structure is described here for the first time in the atmosphere.

The ECMWF analysis uses climatological aerosol fields, so it does not take the aerosol emissions by the Australian wildfires into account, nor the satellite measurements of aerosol extinction. Thus, the replication of the vortex by the analysis was due to assimilation of temperature, wind and ozone measurements from operational satellites. We found that the radio-occultation temperature profiling satellite constellation played the key role in the successful reconstruction of this unusual stratospheric phenomenon by ECMWF analyses (Supplementary notes 4.2). The analyses were also influenced by a few radiosonde profiles and wind profiles in the lower stratosphere by the ESA Aeolus satellite's Doppler wind lidar[28], which provided an observational evidence of the anticyclonic vortex.

The ECMWF IFS produce analyses using a 4-dimensional variational data assimilation method[29], combined with a high-resolution time-evolving forecast model that predicts the atmospheric dynamics. The assimilation system updates the state of the atmosphere using satellite data and in situ observations. As aerosols are not assimilated in the IFS, the ECMWF forecast was not expected to maintain the vortex as observed. Indeed, as shown in Fig. 7a,b, the forecast consistently predicts a decay of the vortex amplitude and fails to predict its rise except on 3 March where the vorticity centroid underwent a jump following the stretching and breaking of the vortex under the effect of vertical shear a few days earlier. As the nature of this event was purely dynamical, it was correctly predicted. This trend would lead to a rapid loss of the vortex and is corrected by the assimilation of new observations, providing an additional forcing that ensures maintenance and rise of the structure.

The ozone hole is forced by the assimilation of satellite observations of ozone (Supplementary Figure 6b,f). The physical mechanism leading to the ozone hole must be a combination of the uplift of ozone-poor tropospheric air and ozone-depleting chemistry in the smoke cloud.

The generation of the smoke-charged vortex is reported in another study[30], of which we were made aware during the review process. They use MLS, OMPS and CALIOP satellite instruments as well as the US Navy Global Environmental Model (NAVGEM) analysis to characterize the chemical composition, thermal structure and the spatiotemporal evolution of the main vortex. They show the evolution of the vortex until 10$^{th}$ March 2020, that is three weeks before it has reached its apogee and collapsed, as follows from our analysis. In contrast to the respective study[30], we provide a more comprehensive analysis of the smoke vortices, quantify the large-scale perturbation of the stratospheric composition and radiative balance, and describe the impact of satellite data assimilation. While the ref. 30 reveals a number of remarkable similarities with respect to our approach and analysis, they report the average diabatic ascent rate of 8 K day$^{-1}$ as opposed to 5.9 K day$^{-1}$ reported here. An important point regarding the vortex dynamics and maintenance, which is left without mention in ref. 30, is the need of a mechanism to suppress the negative potential vorticity produced by the isolated heating as discussed hereinbefore.

## Conclusions

The observed and modeled planetary-scale repercussions of the Australian PyroCb outbreak around the turn of 2020 revolutionize the current understanding and recognition of the climate-altering potential of the wildfires. A single stratospheric overshoot of combustion products produced by the New Year's eve PyroCb event has led to an unprecedented hemispheric-scale perturbation of the stratospheric gaseous and aerosol composition, radiative balance and dynamical circulation with a prolonged effect. Whilst rivaling the volcanic eruptions in terms of stratospheric aerosol load perturbation, this exceptionally strong wildfire event had a substantial impact on a number of other climate-driving stratospheric variables such as water vapour, carbon monoxide and ozone. As the frequency and intensity of the Australian wildfires is expected to increase in the changing climate[31], it is possible that this type of extraordinary event will occur again in the future eventually becoming a significant contributor to the global stratospheric composition.

This work reports the self-organization of an absorbing smoke cloud as a persistent coherent bubble coupled with a vortex that produces confinement preserving the compactness of the cloud. This structure is maintained and rises in the calm summer stratosphere due to its internal heating by solar absorption. The intensity, the duration and the extended vertical and horizontal path of this event certainly ranks it as extraordinary. More detailed studies will be necessary to understand fully the accompanying dynamical processes, in particular how a single sign vorticity structure emerges as a response to heating. Whether stratospheric smoke vortices have already occurred during previous large forest fires is to be explored.

**Methods**
- **OMPS-LP**: The Ozone Mapping and Profiler Suite Limb Profiler (OMPS-LP) on the Suomi National Polar-orbiting Partnership (Suomi-NPP) satellite, which has been in operation since April 2012, measures vertical images of limb scattered sunlight[32]. Aerosol extinction coefficient and ozone number density profiles are retrieved from the limb radiance using a two-dimensional tomographic inversion[13] and a forward model that accounts for multiple scattering developed at the University of Saskatchewan[33]. This OMPS-LP USask aerosol product is retrieved at 746 nm and has a vertical resolution of 1-2 km throughout the stratosphere. The aerosol extinction profiles are exploited to analyze the spatiotemporal evolution of the smoke plumes and for computing the mass of particulate matter lofted above the 380 K potential temperature level corresponding to ~12-17 km altitude in the extratropics, see Fig. 2). The aerosol mass is derived from the aerosol extinction data assuming a particle mass extinction coefficient of 4.5 $m^2\ g^{-1}$ (ref. 2).
- **SAGEIII:** The Stratospheric Aerosol and Gas Experiment (SAGE) III provides stratospheric aerosol extinction coefficient profiles using solar occultation observations from the International Space Station (ISS)[34]. These measurements, available since February 2017, are provided for nine wavelength bands from 385–1,550 nm and have a vertical resolution of approximately 0.7 km. The SAGE III/ISS instrument and the data products have characteristics nearly identical to those from the SAGE III Meteor mission[35]. Here we use the 754 nm wavelength band for quantifying the error of OMPS-LP aerosol extinction retrieval (using 16-84 percentiles) and the wavelength pair 1019/521 nm for deriving the Angstrom exponent, which is used for the radiative forcing calculations.
- **CALIOP**: The Cloud-Aerosol Lidar with Orthogonal Polarization (CALIOP) is a two-wavelength polarization lidar on board the CALIPSO mission[10] that performs global profiling of aerosols and clouds in the troposphere and lower stratosphere. We use the total attenuated 532 nm backscatter level 1 product V3.40 which is available in near real time with a delay of a few days

(doi:10.5067/CALIOP/CALIPSO/CAL_LID_L1-VALSTAGE1-V3-40). The along track horizontal / vertical resolution are respectively 1 km / 60 m between 8.5 and 20.1 km, 1.667 km / 180 m between 20.1 and 30.1 km, and 5 km / 300 m resolution between 30.1 km and 40 km. The L1 product oversamples these layers with an actual uniform horizontal resolution of 333 m.

- **MLS**: The MLS (Microwave Limb Sounder)[16] instrument on the NASA Aura satellite has been measuring the thermal microwave emission from Earth's atmospheric limb since July 2004. With ~15 orbits per day, MLS provides day and night near-global (82°S - 82°N) measurement of vertical profiles of various atmospheric gaseous compounds (including $H_2O$, CO and $CH_3CN$) cloud ice, geopotential height, and temperature of the atmosphere. The measurements yield around to 3500 profiles per day for each species with a vertical resolution of ~3 - 5 km. For tracking the smoke bubble, we selected profiles bearing CO enhancements in the stratosphere exceeding 400 ppbv and/or $H_2O$ enhancement exceeding 12 ppmv with respect to the pre-event conditions (late December 2020). Stratospheric mass loads of $H_2O$, CO and $CH_3CN$ are derived from MLS volume mixing ratio measurements of species in log pressure space, molecular mass of the compound and the air number density derived from MLS temperature profile on pressure levels above the 380 K isentropic level and between 20°S and 82°S. The error bars on the mass of injection are estimated by combining accuracies on the measurements and the mean standard deviations over 20-day periods before and after the sharp increase. The error bar on the $CH_3CN$ mass above 380K is only calculated on the standard deviation because accuracies on $CH_3CN$ measurements are extremely large. The error bar on the aerosol mass takes also into account the uncertainty on the particle mass extinction coefficient (1.5 $m^2g^{-1}$).

- **TROPOMI**: The TROPOspheric Monitoring Instrument (TROPOMI) is a nadir viewing shortwave spectrometer jointly developed by the Netherlands Space Office and the European Space Agency on board the Sentinel 5 Precursor mission. The instrument has spectral bands in the ultraviolet (270 – 500 nm), the near infrared (710 – 770 nm) and the shortwave infrared (2314 – 2382 nm) providing therefore observation of key atmospheric constituents, among which we use $O_3$, CO and aerosol index at high spatial resolution (7×3.5 $km^2$ at nadir for the UV, visible and near-infrared bands, 7×7 $km^2$ at nadir for shortwave infrared bands)[36]. Here we exploit the Aerosol Index, and the CO and $O_3$ columnar values from the offline Level 2 data. The Aerosol Absorbing Index (AI) is a quantity based on the spectral contrast between a given pair of UV wavelength (in our case the 340--380 nm wavelengths couple). The retrieval is based on the ratio of the measured top of the atmosphere reflectance (for the shortest wavelength) and a pre-calculated theoretical reflectance for a Rayleigh scattering-only atmosphere (assumed equal for both wavelengths). When the residual value between the observed and modelled values is positive, it indicates the presence of UV-absorbing aerosols, like dust and smoke. Negative residual values may indicate the presence of non-absorbing aerosols, while values close to zero are found in the presence of clouds. AI is dependent upon aerosol layer characteristics such as the aerosol optical thickness, the aerosol single scattering albedo, the aerosol layer height and the underlying surface albedo. Providing a daily global coverage and a very high spatial resolution (7×3.5 $km^2$ at the nadir), the AI from TROPOMI is ideal to follow the evolution of smoke, dust, volcanic ash, or aerosol plumes.
Ozone:
Copernicus Sentinel-5P (processed by ESA), 2018, TROPOMI Level 2 Ozone Total Column products. Version 01. European Space Agency. https://doi.org/10.5270/S5P-fqouvyz
Carbon Monoxide:
Copernicus Sentinel-5P (processed by ESA), 2018, TROPOMILevel 2 Carbon Monoxide total column products. Version 01. European Space Agency. https://doi.org/10.5270/S5P-1hkp7rp
Aerosol Index:
Copernicus Sentinel-5P (processed by ESA), 2018, TROPOMILevel 2 Ultraviolet Aerosol Index products. Version 01. European Space Agency. https://doi.org/10.5270/S5P-0wafvaf

- **GNSS-RO**: We use Global Navigation Satellite System (GNSS) Radio Occultation (RO) dry temperature profiles acquired onboard Metop A/B/C satellites and processed in near real time mode at EUMETSAT RO Meteorology Satellite Application Facility (ROM SAF)[37]. For computing the composited temperature perturbation within the smoke bubble we use temperature profiles collocated with the vortex centroid as identified using IFS analyses (8 hours, 400 km collocation criteria). The perturbation is computed as departure from a mean temperature profile within the corresponding spatiotemporal bin (3-day, 3° latitude, 40° longitude).

- **ECMWF IFS** is the operational configuration of the ECMWF global Numerical Weather Prediction system (46R1, https://www.ecmwf.int/en/publications/ifs-documentation). It consists of an atmosphere-land-wave-ocean forecast model and an analysis system that provides an accurate estimate of the initial state. The forecast model has a 9 km horizontal resolution grid and 137 vertical levels, with a top around 80 km altitude. The analysis is based on a 4-dimensional variational method, run twice daily using more than 25 million observations per cycle, primarily from satellites. The IFS produces high-resolution operational 10-day forecasts twice daily.

- **Radiative transfer calculations:** The equinox-equivalent daily-average shortwave (integrated between 300 and 3000 nm) surface and top of the atmosphere (TOA) direct radiative forcing (RF) are estimated using the UVSPEC (UltraViolet SPECtrum) radiative transfer model in the libRadtran (library for Radiative transfer) implementation[38] and a similar methodology as in ref. 39 and ref. 4. Baseline and fire-perturbed simulations are carried out with different aerosol layers: the average OMPS-LP aerosols extinction coefficient profiles, for January and February 2019 (baseline simulation) and January and February 2020 (fire-perturbed simulation). The spectral variability of the aerosol extinction is modeled using the measured Ångström exponent from SAGE III for January 2020 (fire-perturbed simulation) and typical background values inferred from SAGE III (baseline simulation). Different hypotheses have been considered for the non-measured optical parameters of fire aerosols: single scattering albedo from 0.85 to 0.95 (typical of wildfire aerosols, see e.g. ref. 40) and a Heyney-Greenstein phase function with an asymmetry parameter of 0.70. More information about the UVSPEC runs can be found in the Supplementary Material. The daily-average shortwave TOA radiative forcing for the fire-perturbed aerosol layer is calculated as the SZA-averaged upward diffuse irradiance for a baseline simulation without the investigated aerosols minus that with aerosols, integrated over the whole shortwave spectral range. The shortwave surface radiative forcing is calculated as the SZA-average downward global (direct plus diffuse) irradiance with aerosols minus the baseline, integrated over the whole spectral range.

- **Mass estimation**

    The stratospheric masses of CO, $CH_3CN$, and $H_2O$ are derived from the MLS species mixing ratios vertical profiles combined with MLS vertical profiles of pressure and temperature on the standard 37 pressure levels. MLS data are filtered according to the recommendations from the MLS team (https://mls.jpl.nasa.gov/data/v4-2_data_quality_document.pdf). $CH_3CN$ measurements are not recommended below 46 hPa however have already been used successfully in a study on combustion products from Australian bush fires in the stratosphere[41]. The mass calculation is performed as follows. First, we calculate for each profile the partial column of species (CO, $CH_3CN$, $H_2O$) and air for each measurement layer (vertical resolution of ~3 km); the air partial column is derived from the difference in pressure between the top and bottom of the layer. Then, adding the partial columns in a profile, we derive the total column of species and air above the 380 K potential temperature (~ 12 .. 17 km altitude), the so-called stratospheric "overworld"[14]. The ratio of the species and air total columns gives us the mean volume mixing ratios (VMR) of the species over the stratospheric profile. Finally, we calculate the mean VMR of all the profiles between 20 °S and 82 °S (which is the southernmost latitude of the MLS sampling)

and multiply it by the molecular mass of the species and the total number of air molecules above 380 K and between 20 °S and 82 °S following ref. 42 to obtain the total mass burden of the species plotted in Fig. 4. The aerosol mass is derived from the OMPS satellite aerosol extinction data assuming a particle mass extinction coefficient of 4.5 m$^2$g$^{-1}$ following ref. 2. Standard deviations of the injected mass are estimated by combining accuracies on the measurements and the mean standard deviations over 20-day periods before and after the sharp increase. The standard deviation of the $CH_3CN$ mass above 380 K is only calculated from the standard deviation because accuracies on $CH_3CN$ measurements are extremely large. The standard deviation of the aerosol mass takes also into account the uncertainty on the particle mass extinction coefficient (error=1.5 m$^2$g$^{-1}$).

## Data availability

MLS data are publicly available at http://disc.sci.gsfc.nasa.gov/Aura/data-holdings/MLS; GNSS-RO data at https://www.romsaf.org/product_archive.php; OMPS-LP data at ftp://odin-osiris.usask.ca/ with login/password osirislevel2user/hugin ; SAGE III data at doi:10.5067/ISS/SAGEIII/SOLAR_BINARY_L2-V5.1 ; CALIOP data at doi:10.5067/CALIOP/CALIPSO/CAL_LID_L1-VALSTAGE1-V3-40 ; TROPOMI ozone data at https://doi.org/10.5270/S5P-fqouvyz ; carbon monoxide data at https://doi.org/10.5270/S5P-1hkp7rp ; Aerosol Index data at https://doi.org/10.5270/S5P-0wafvaf. The extracted ECMWF data used in this work are available at https://doi.org/10.5281/zenodo.3958214

## Code availability

LibRadTran code exploited for radiative forcing calculations is available at http://www.libradtran.org/doku.php?id=download

The processing code for CALIOP, ECMWF data is available at https://github.com/bernard-legras/STC-Australia with dependencies in https://github.com/bernard-legras/STC/tree/master/pylib.

The processing code for TRIPOMI is available at https://github.com/silviabucci/TROPOMI-routines

The processing codes for MLS, OMPS-LP, SAGEIII, GNSS-RO are available at https://doi.org/10.5281/zenodo.3959259.

The processing code for the estimation of injected masses using MLS data is available at https://doi.org/10.5281/zenodo.3959350

## Acknowledgements

CALIOP data were provided by the ICARE/AERIS data centre. The TROPOMI data were provided by the Copernicus Open Access Web https://scihub.copernicus.eu/. We thank the EUMETSAT's Radio Occultation Meteorology Satellite Application Facility (ROM SAF) for providing NRT temperature profile data . MLS data are provided by the NASA Goddard Space Flight Center Earth Sciences (GES) Data and Information Services Center (DISC). We thank the OMPS-LP team at NASA Goddard for producing and distributing high quality Level 1 radiances, and the SAGE III/ISS team at NASA Langley for data production and advice, in particular Dave Flittner. We acknowledge the support of ANR grant 17-CE01-0015. The providers of the libRadtran suite (http://www.libradtran.org/) are gratefully acknowledged. We acknowledge discussions with Guillaume Lapeyre, Riwal Plougonven and Aurélien Podglajen.

**Supplementary Information** is available for this paper.

The authors declare no competing interests.

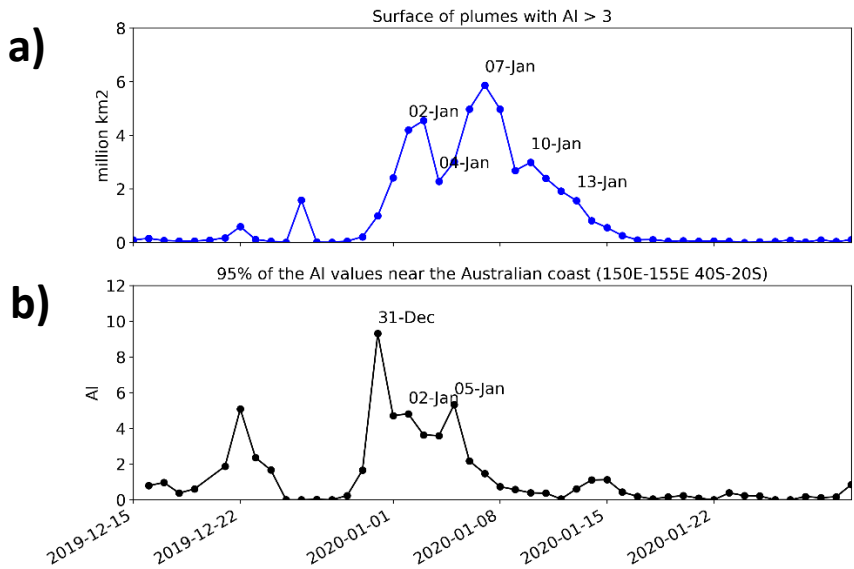

**Figure 1. Time evolution of the smoke clouds as observed by TROPOMI satellite instrument.** a) time evolution of the total surface covered by the aerosol plumes with Absorbing Aerosol Index AAI>3 over the southern hemisphere. This threshold is chosen to follow the evolution of the main plume that was characterized by values of AI up to 10. The plumes show a sharp gradient in AI at the borders, where the AI value rapidly decreases, allowing to clearly define the boundaries of the aerosol cloud. b) 95th percentile of the aerosol close to the Eastern Australian coastal region (150°-155 E 20°-40° S) where the extreme PyroCb activity took place. The main aerosol injections occurred between 30 and 31 December 2019, producing a plume that reached a first maximal spatial extension on the 2 January 2020, and between 4 and 5 January, when a second event produced an additional aerosol cloud that, combined with the first one, caused a total absorbing aerosol coverage that reached a maximum of 6 millions km$^2$ of extension on the 7 January. The plumes then gradually dissipated and diluted, decreasing in their AI values, until the third week of January, when the AI signal from the aerosol clouds is no more visible by TROPOMI, with the exception of few bubbles of confined aerosol (see next sections)

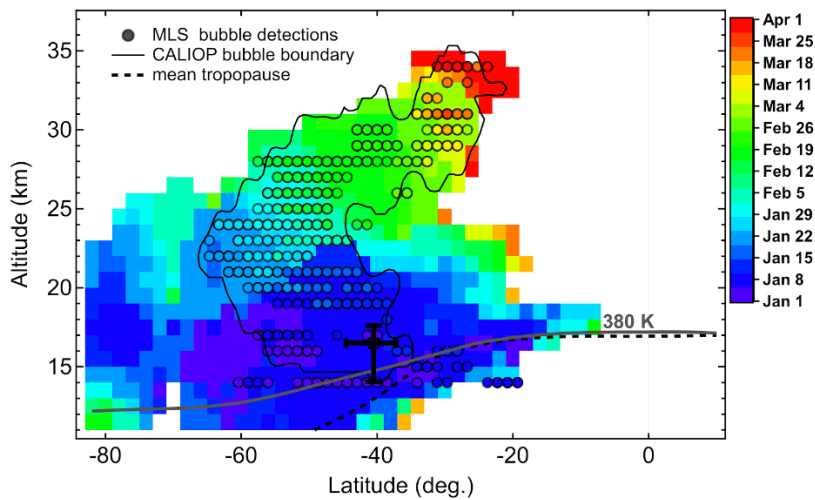

**Figure 2. Latitude-altitude evolution of the smoke plumes in the stratosphere.** The pixels, colour coded by date, indicate doubling of aerosol extinction with respect to December 2019 levels for data where aerosol to molecular extinction ratio is 1 or higher. The black circles with date-colour filling indicate the locations of high amounts of water vapour and/or carbon monoxide detected by MLS (see Methods). The black contour encircles the locations of aerosol bubble detections by CALIOP lidar (see Fig. 5a and Supplementary Figure 3). The cross marks the latitude-altitude extent of the stratospheric cloud detected by CALIOP on the 1st January (see Fig 5a). The grey solid and black dashed curves indicate respectively the zonal-mean 380 K isentrope and the lapse rate tropopause for the January-March 2020 period.

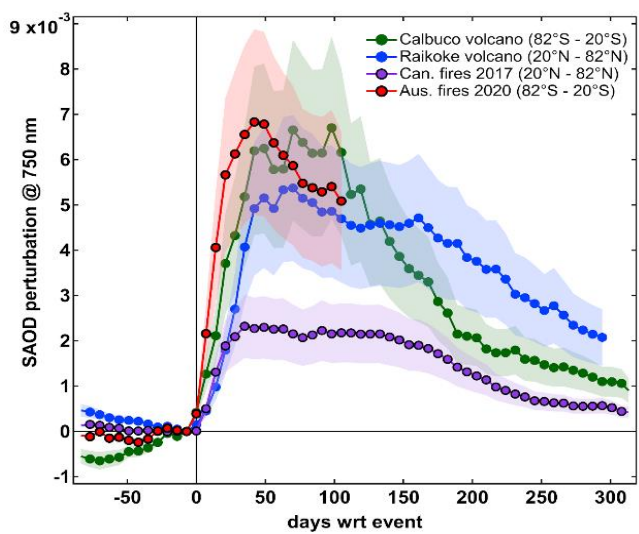

**Figure 3. Perturbation of the stratospheric aerosol optical depth (SAOD) due to Australian fires and the strongest events since 1991**. The curves represent the SAOD perturbation at 746 nm following the Australian wildfires, the previous record-breaking Canadian wildfires in 2017 and the strongest volcanic eruptions in the last 29 years (Calbuco, 2015 and Raikoke, 2019 [ref. 11]). The time series are computed from OMPS-LP aerosol extinction profiles as weekly-mean departures of aerosol optical depth above 380 K isentropic level (see Fig. 2) from the levels on the week preceding the event. The weekly averages are computed over equivalent-area latitude bands (as indicated in the panel) roughly corresponding to the meridional extent of stratospheric aerosol perturbation for each event. The shading indicates a 30% uncertainty in the calculated SAOD, as estimated from SAGE III coincident comparisons (See Methods).

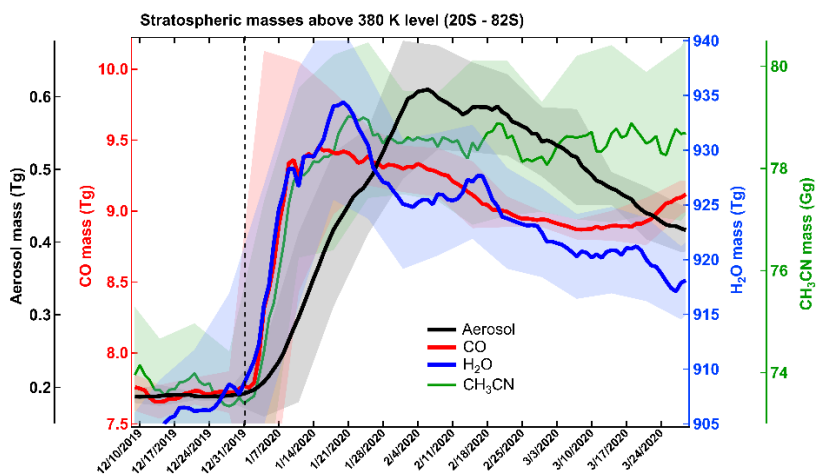

**Figure 4. Time evolution of the daily total mass of CO, CH$_3$CN, H$_2$O and aerosols above the 380 K potential temperature, between 20 °S and 82 °S.** The dotted and solid lines correspond to daily data and 1-week smoothed data respectively. Envelopes represent two standard deviations over the 1-week window (See Methods). As shown in this figure, the levels of CO, CH$_3$CN, H$_2$O, and aerosols started to increase simultaneously and kept increasing during ~2-3 weeks, a duration corresponding probably to the time taken by products injected in the lowermost stratosphere to ascend above 380 K. The stratospheric masses of carbon monoxide (CO) and acetonitrile (CH$_3$CN) bounded within the southern extratropics increase abruptly by 1.5±0.9 Tg and 3.7±2.0 Gg respectively during the first week of 2020. This gives a CO/CH$_3$CN mass ratio of 0.0025, consistently with previous estimates for temperate Australian wildfires[15]. The injected mass of water was estimated at 27±10 Tg that is about 3% of the total mass of stratospheric overworld water vapour in the southern extratropics. The shading shows that the amplitude of fluctuations increases sharply during the sharp rise of species masses, reflecting the fact that sampling of the bubble by MLS is more random than on a more homogeneous field. The lagging increase of the aerosol mass is due to the fact that the OMPS-LP extinction retrieval saturates at extinction values above 0.01 km$^{-1}$. Profiles are therefore truncated below any altitude exceeding this value, which can lead to an underestimation of the early aerosol plume when it is at its thickest. This artifact, which explains the slower increase of aerosol mass than gases, persists until mid-February when the plume is sufficiently dispersed so that OMPS-LP extinction measurements no longer saturate.

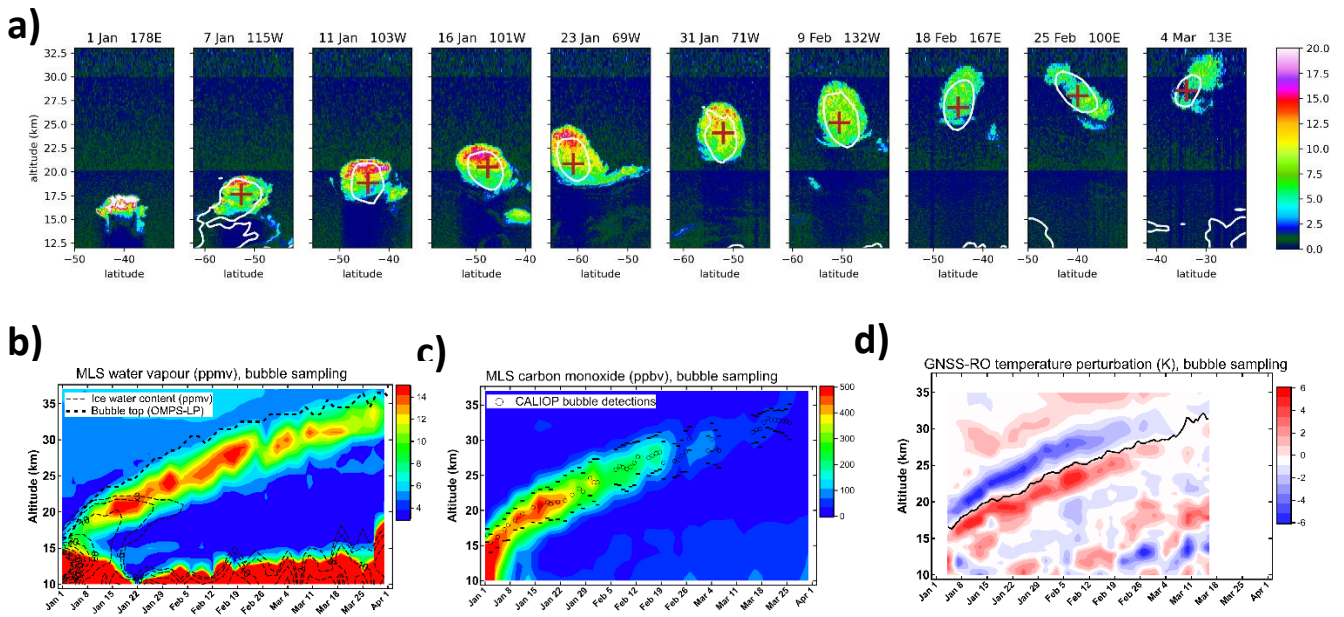

**Figure 5: Vertical evolution of the smoke bubble, its chemical composition and thermal structure**.
a) Selection of CALIOP attenuated scattering ratio profiles for clear intersections of the bubble by the orbit except the first panel of 1st of January that shows the dense and compact plume on its first day in the stratosphere. The attenuated scattering ratio is calculated by dividing the attenuated backscattering coefficient by the calculated molecular backscattering. The data are further filtered horizontally by an 81-pixel moving median filter to remove the noise. The crosses show the projected interpolated location of the vortex vorticity centroid from the ECMWF operational analysis onto the orbit plane at the same time and the white contour shows the projected contour of the half maximum vorticity value in the pane passing by the vortex centroid and parallel to the orbit plane (See Fig. 6). b) Evolution of the water vapour mixing ratio within the rising bubble based on MLS bubble detections (see Methods). The dashed contours show the equivalent mixing ratio of ice water derived from MLS ice water content vertical profiles collocated with the bubble. The thick dashed curve marks the top altitude of the aerosol bubble determined as the level where OMPS-LP extinction triples that of the nearest upper altitude level. c) Evolution of carbon monoxide (MLS) within the bubble. The centroid and the vertical boundaries of the aerosol bubble determined using CALIOP data are overplotted as circles and bars respectively. d) Composited temperature perturbation within the smoke bubble from Metop GNSS radio occultation (RO) temperature profiles collocated with the smoke bubble (see Methods). The black line shows the centroid of vortex detected from ECMWF data (see Supplementary notes 4.1 and Supplementary Figure 7b).

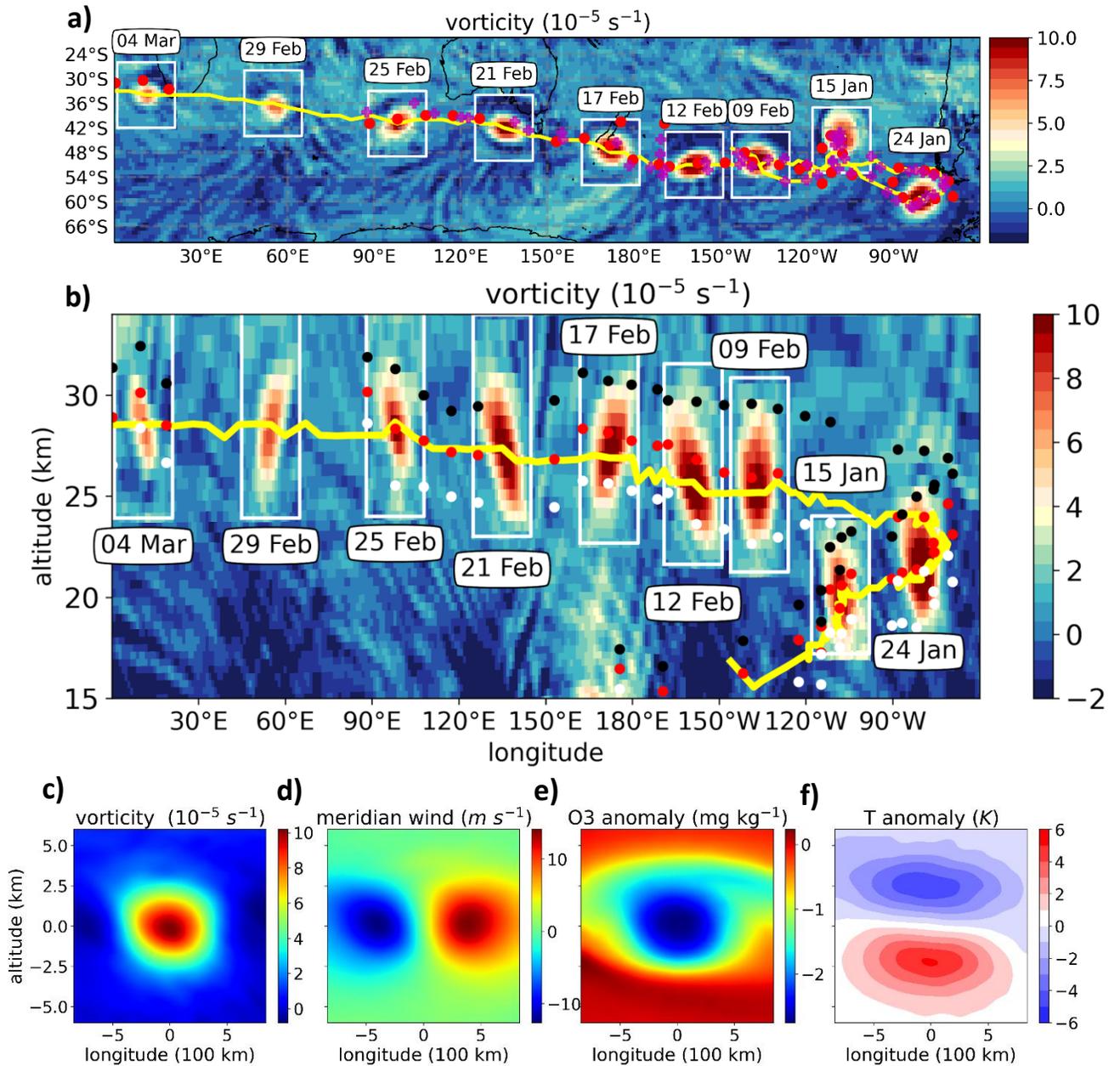

**Figure 6: Spatiotemporal evolution of the vortex and its thermodynamical properties.** a) Composite horizontal sections of the vortex. The background shows the relative vorticity field on 24 Jan 2020 6UTC from the ECMWF operational analysis on the surface 46.5 hPa (21.3 km at the location of the vortex) corresponding to the level of highest vorticity in the vortex. The boxes show the vorticity field at other times as horizontal sections at the level of maximum vorticity centroid projected onto the background field. The yellow curve is the twice-daily sampled trajectory of the vortex centroid. The red dots show the location of the CALIOP bubble centroid for all the cases where it is clearly intersected by the orbit. The magenta crosses show the location of the center of the compact aerosol index anomaly as seen from TROPOMI (Supplementary notes 3).

b) Composite vertical section of the vortex. The background is here the longitude-altitude section of the vortex on 24 Jan 2020 6 UTC at the latitude 47°S. The boxes show vertical sections at the same time as panel a) at the latitude of maximum vorticity. The black, red and white dots show, respectively, the CALIOP bubble top, centroid and bottom.

c) Composite of the vortex vorticity in the longitude altitude plane at the level and at the latitude of the vortex centroid performed during the most active period of the vortex between 14 Jan and 22 February.

d) Same as c) for the meridional wind deviation with respect to the mean in the displayed box.

e) Same as c) but for the ozone mixing ratio deviation with respect to the zonal mean.

f) Same as c) but for the temperature deviation with respect to the zonal mean.

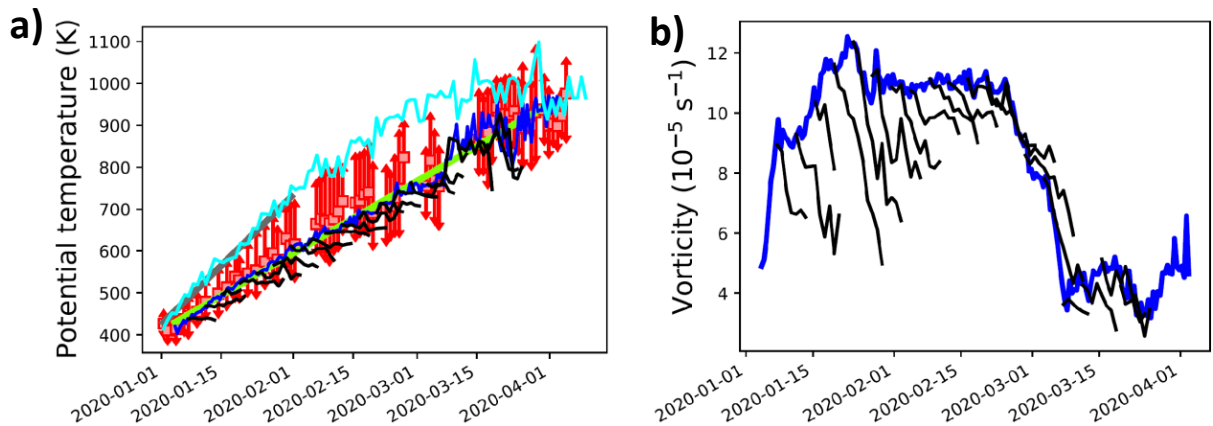

**Figure 7: Time evolution of the altitude and vorticity of the main vortex.** The two panels show the potential temperature (a) and vorticity (b) as a function of time. All the quantities are defined at the vortex centroid where the vorticity is maximum. In panel a) the red squares show the position of the aerosol bubble centroid according to CALIOP. The CALIOP centroid is defined by averaging the most extreme top, bottom, south and north edges. The arrows show the extension of the bubble in potential temperature space. The cyan line shows the upper envelope of the bubble as detected by OMPS-LP, the green line is a linear fit to the ascent of the vortex and the gray line shows a 10 K day$^{-1}$ curve. In panels (a) and (b), the black lines show the ECMWF 10-day forecast evolution, plotted every four days. The forecast evolution is only shown for the period where it maintains the vortex. The slight discrepancy between the analysis and the initial point of the forecast is because the 10-day forecast is produced from a slightly inferior 6-hour analysis, due to real time constraints.

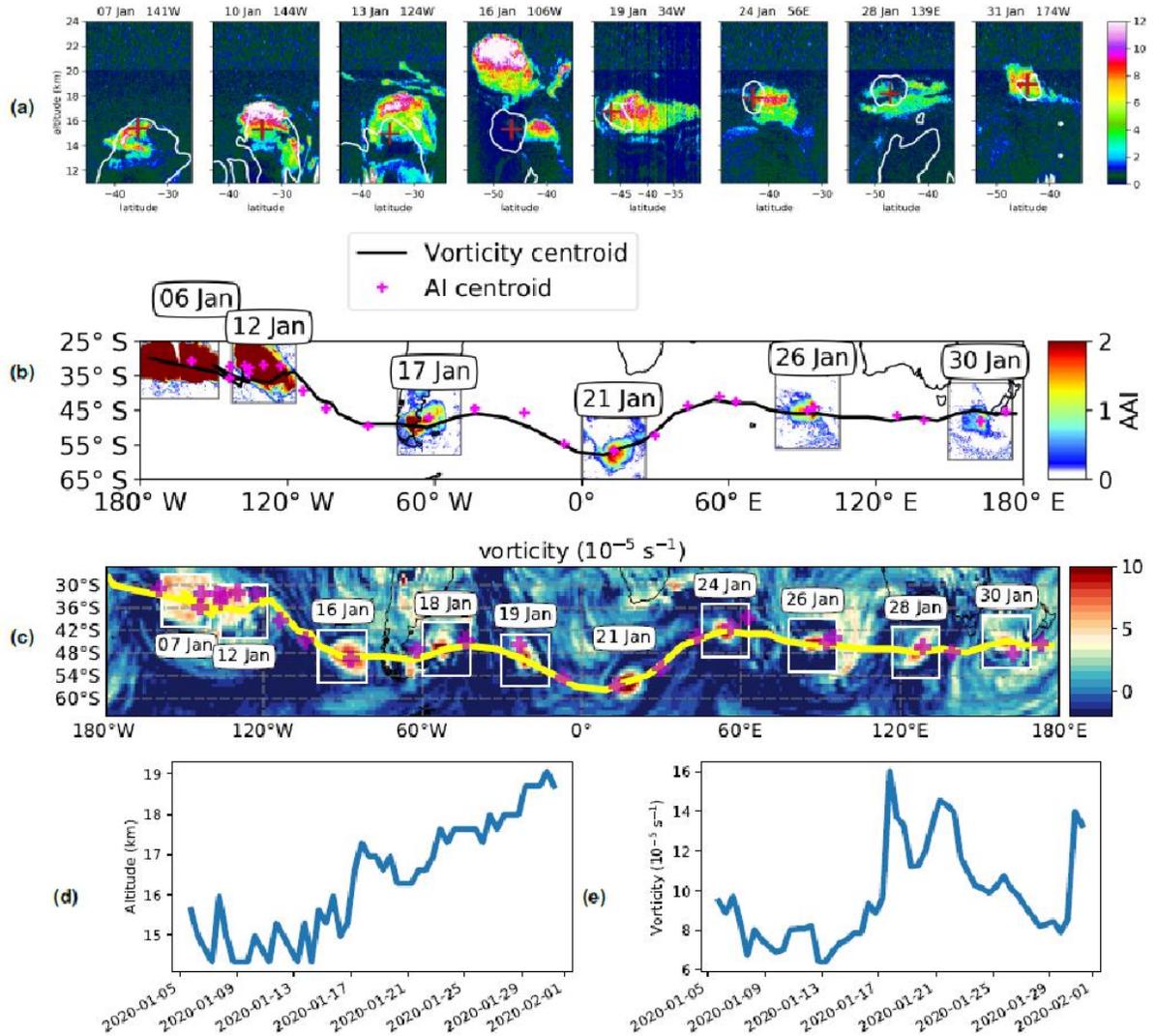

**Figure 8: Spatiotemporal evolution of the second vortex**. (a) Time evolution from eight matching sections of CALIOP. (b) Composite of TROPOMI Aerosol Index (AI) at the location of the vortex for six dates that do not necessarily match that of CALIOP. (c) Time evolution of the vortex according the ECMWF analysis from ten vorticity snapshots at the level of maximum vorticity. The background is shown for 24 January. In b) and c) the trajectory of the AI centroid is shown as magenta crosses and the trajectory of the IFS vortex is shown as the black and yellow curves, respectively. (d) Altitude of the vortex core as a function of time. (e) Maximum vorticity at the vortex core as a function of time.

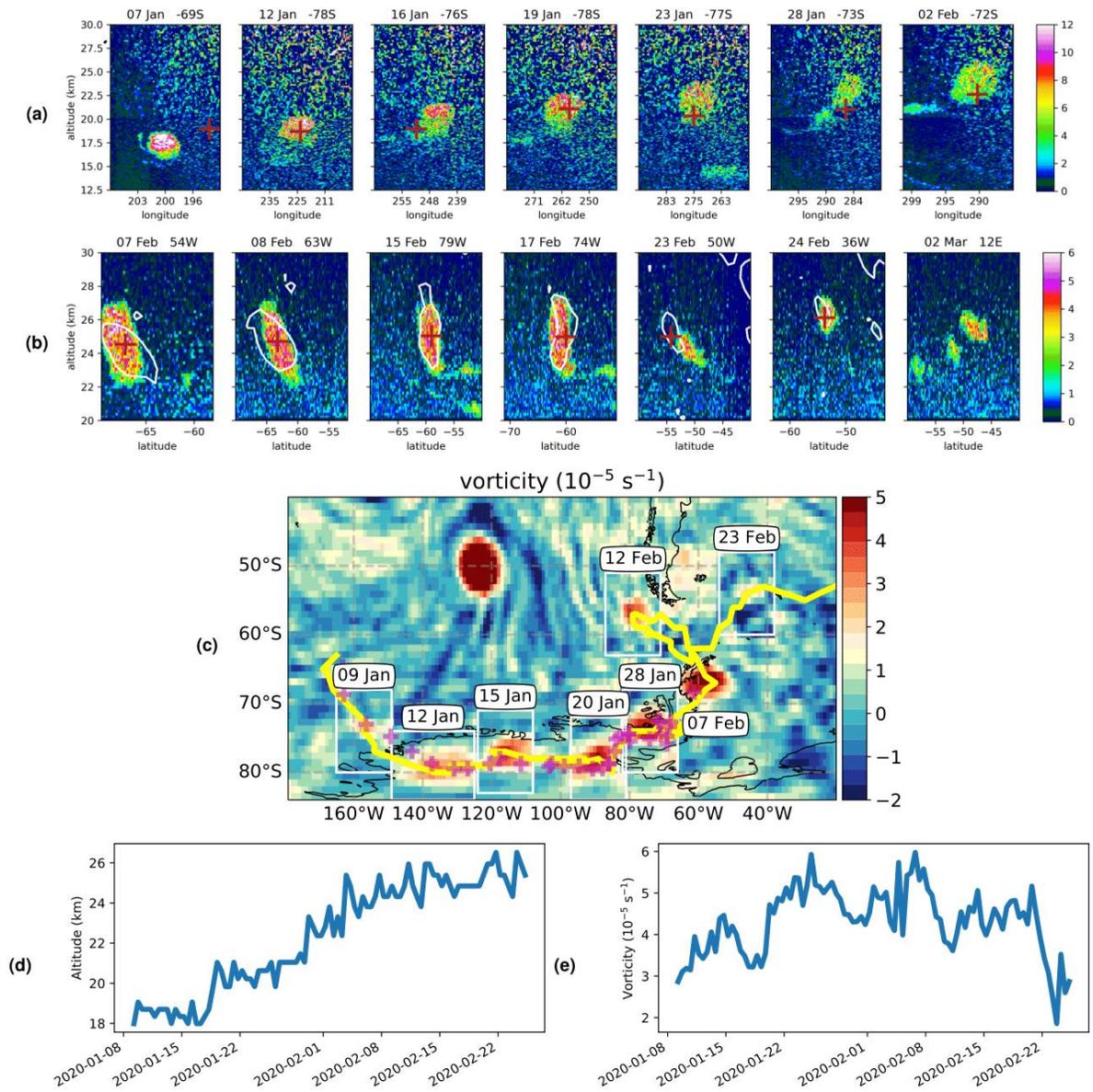

**Figure 9: Spatiotemporal evolution of the third vortex** (a) Time evolution of the vortex from 7 matching sections of CALIOP during its first period over Antarctica. We use here daily orbits of CALIOP, hence the high level of noise. The x-axis is mapped over longitude due to the proximity of the pole. (b) Time evolution of the third vortex from 7 matching sections of CALIOP during its second period. (c) Time evolution of the vortex according the ECMWF analysis from eight vorticity snapshots at the level of maximum vorticity. The background is shown for 7 February where the main vortex is also visible. The yellow curve shows the trajectory of the vortex in the IFS, the magenta crosses mark the TROPOMI AI centroid (d) Altitude of the vortex core as a function of time. (e) Maximum vorticity of the vortex core as a function of time.

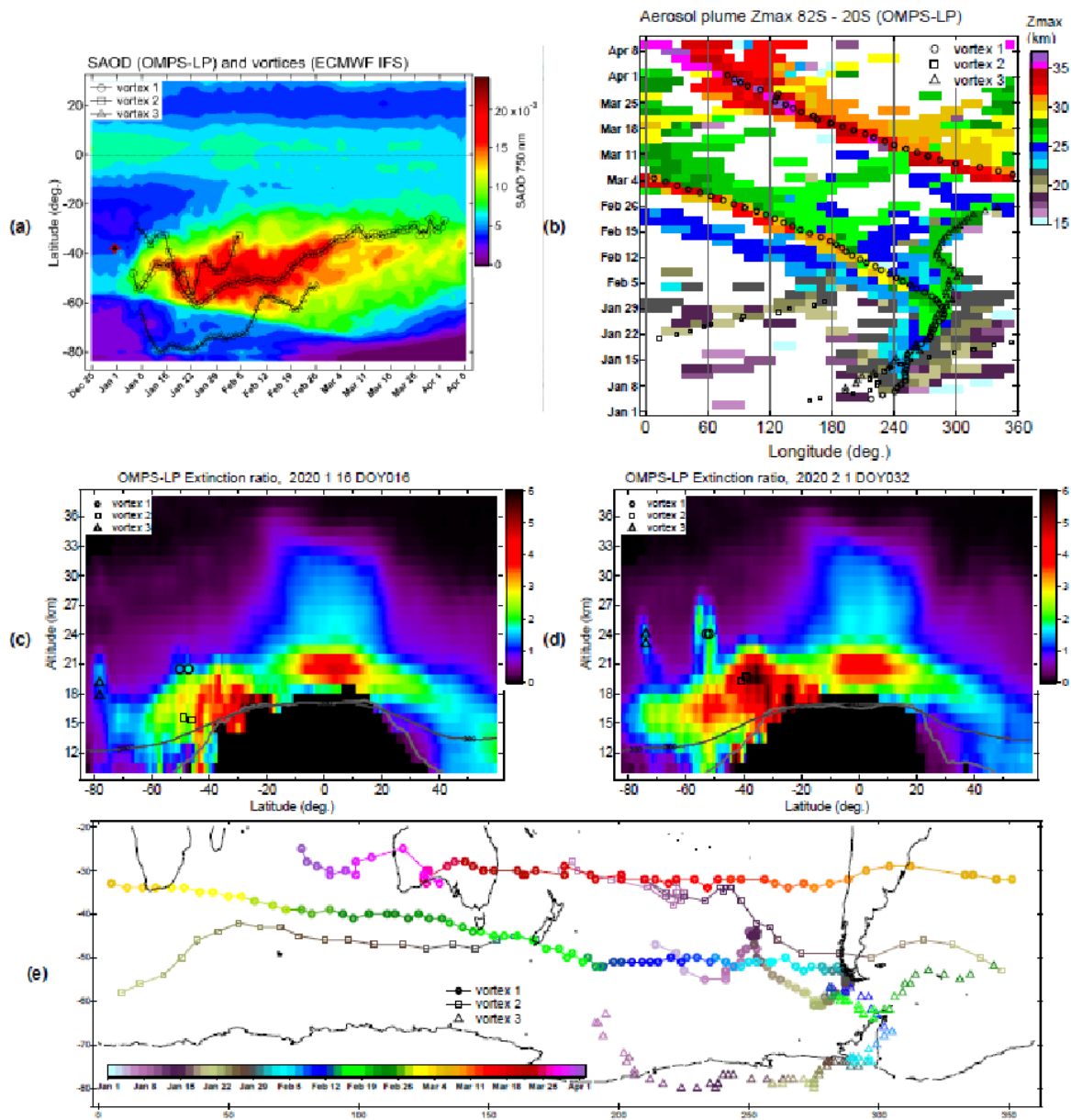

**Figure 10: Three-dimensional evolution of the three vortices from OMPS-LP and ECMWF IFS.**
(a) Time-latitude section of zonal-mean stratospheric aerosol optical depth (above the tropopause) from OMPS-LP measurements. The markers show locations of smoke-charged vortices identified using ECMWF vorticity fields. (b) Longitude-temporal evolution (Hovmöller diagram) of the maximum altitude of smoke plume inferred from OMPS-LP extinction data within 30 °S-60 ° above 15 km where aerosol to molecular extinction ratio exceeds 5. The markers indicate the locations of the smoke-charged vortices identified using ECMWF vorticity fields. (c) Latitude-altitude section of zonal-mean aerosol to molecular extinction ratio above the local tropopause from OMPS-LP measurements for 16 January 2020. The thick and the thin curves indicate, respectfully, the zonal-mean lapse-rate tropopause and the 380 K potential temperature level. The markers show the positions of the main vortex and of the two companion vortices. (d) Same as (c) for 1 February 2020. (e) Trajectories of the three vortices with colour-coded date in the longitude-latitude plane.

# Supplement material to "The 2019/2020 Australian wildfires generated a persistent smoke-charged vortex rising to 35 km altitude"

Sergey Khaykin, Bernard Legras, Silvia Bucci, Pasquale Sellitto, Lars Isaksen, Florent Tence, Slimane Bekki, Adam Bourassa, Landon Rieger, Daniel Zawada, Julien Jumelet & Sophie Godin-Beekmann

29 July 2020

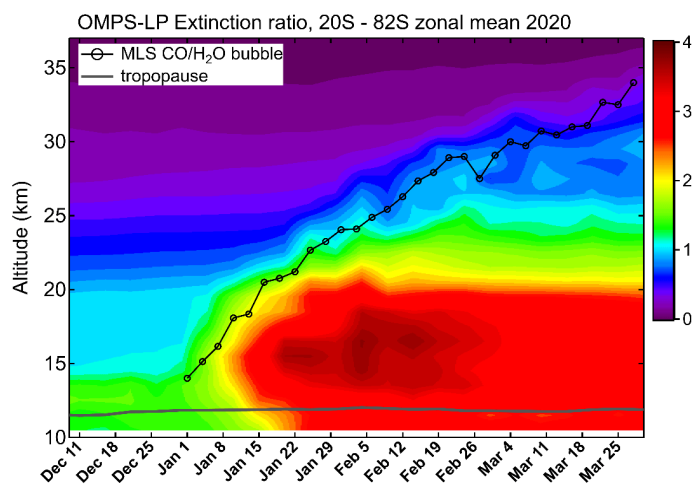

Supplementary Figure 1: **Time evolution of OMPS profile and MLS bubble**
Time evolution of OMPS zonally-averaged extinction ratio profile within 20°S-82S° latitude band. The black curve indicates the centroid altitude of water vapour and/or carbon monoxide enhancements detected by MLS within the rising bubble of smoke (see Methods).

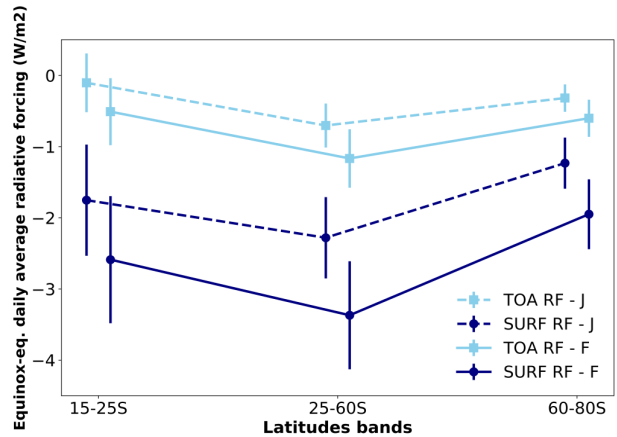

Supplementary Figure 2: **Radiative Forcing**
Equinox-equivalent daily-average regional RF at TOA and surface due to Australian fire perturbations to the stratospheric aerosol layer. Data is averaged over the months of January and February. These estimations are provided for the following regions (latitude bands): 15 to 25°S, 25 to 60°S and 60 to 80°S. The error bars show the variability of our estimations when varying our hypotheses on non-measured quantities (single scattering albedo and asymmetry parameter).

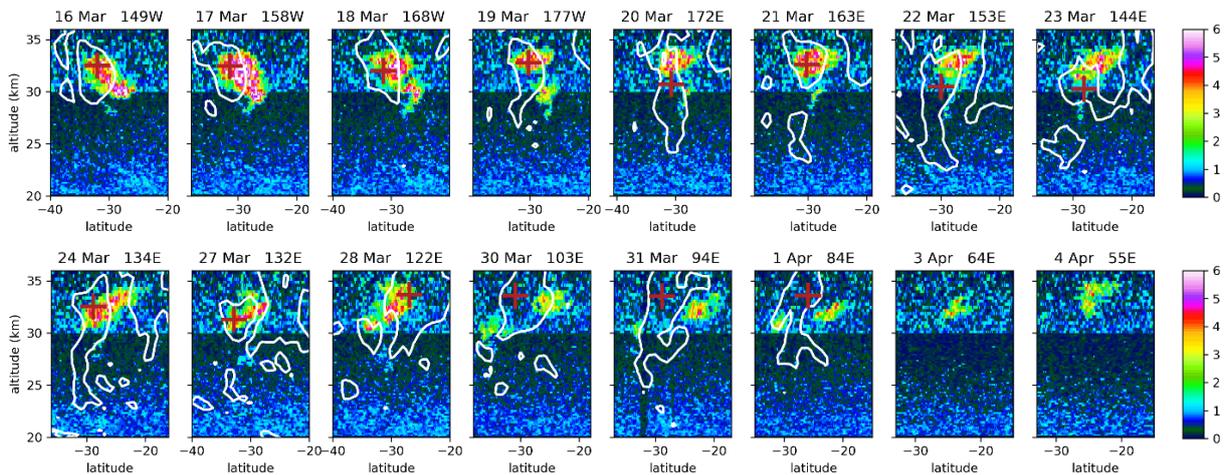

Supplementary Figure 3: **Late evolution of the smoke bubble**
Selection of scattering ratio profiles for intersections of the bubble by CALIOP for orbits between 16 March and 4 April 2020. The shown data are the same as in Figure 2a of the main text except that the median filter is here applied on 161 pixels and the color scale ranges from 0 to 6.

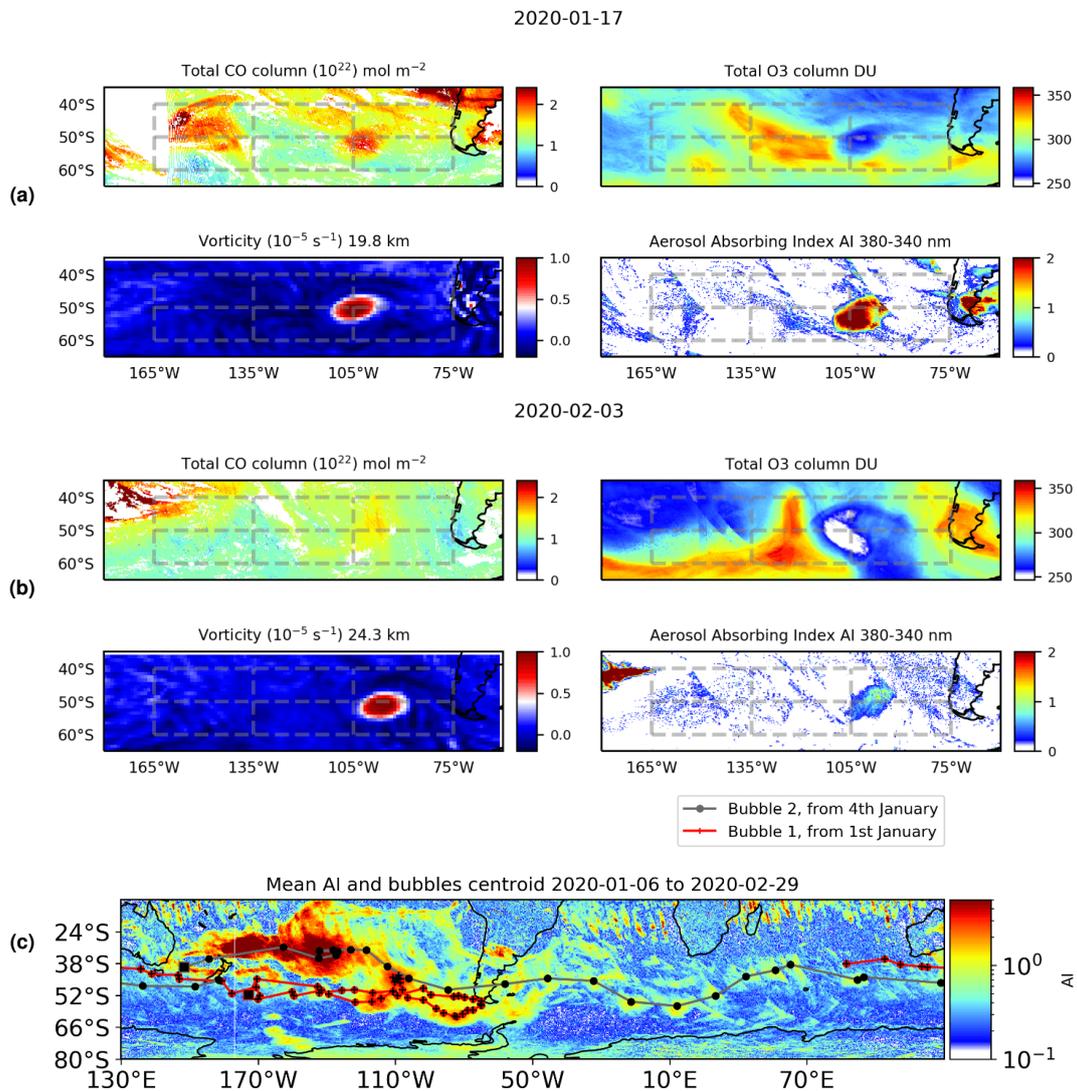

Supplementary Figure 4: **Vortices and tracer confinement as seen from TROPOMI**
Atmospheric composition in correspondence of the main vortex on 17 January (a) and 3 February (b). The 4 panels show the total columnar values of CO concentration (in molecules per m$^2$), the O$_3$ columnar values (in Dobson Units) and the AI, all measured from TROPOMI, as compared to the position of the vorticity anomaly as reproduced by IFS. In (c) is shown the composite of the AI values along the period immediately after the two main fire injection took place (starting from 6 January) and representing therefore the extension of the area concerned by the aerosol perturbation along the period. Overlapped, the red and the black lines represent the position of the centroids of the main and the secondary vortex, respectively, as identified from the AI daily maps. The black dots mark the position every 24 hours for both the main and the secondary vortex from 1 January and 4 January respectively, i.e. the dates at which the plumes where first identified from the AI images.

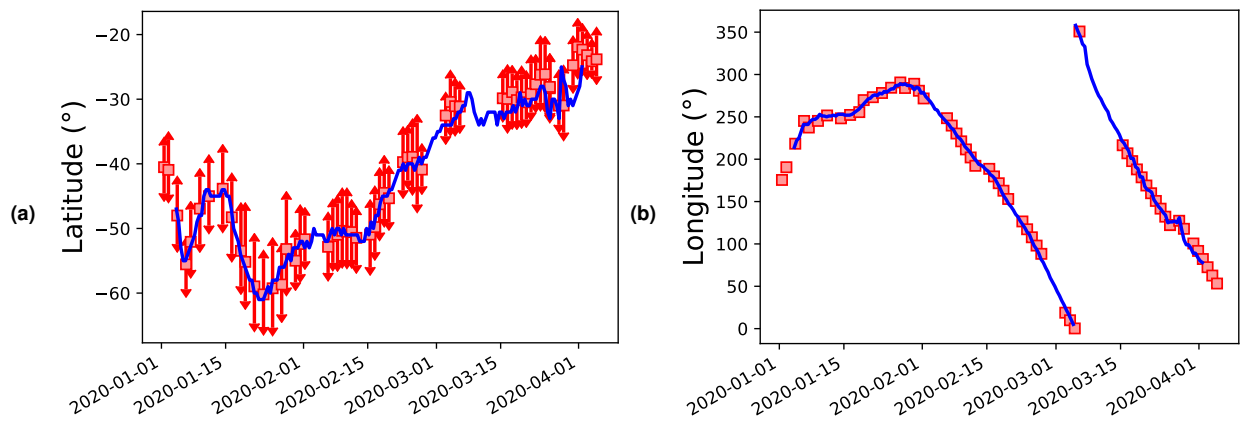

Supplementary Figure 5: **Time evolution of the vortex in the IFS**
Time evolution of the geographical position of the main vortex as seen from the IFS (blue line) and from CALIOP (red squares). (a) latitude, (b) longitude as a function of time. All the quantities are defined at the vortex centroid where the vorticity is maximum. The red squares in the two panels show the position of the aerosol bubble centroid according to CALIOP. The CALIOP centroid is defined by averaging the most extreme top, bottom, south and north edges. The arrows in the latitude and potential temperature panels show the extension of the bubble in those direction. As the orbit almost follows a meridian, the longitude extension cannot be retrieved from CALIOP.

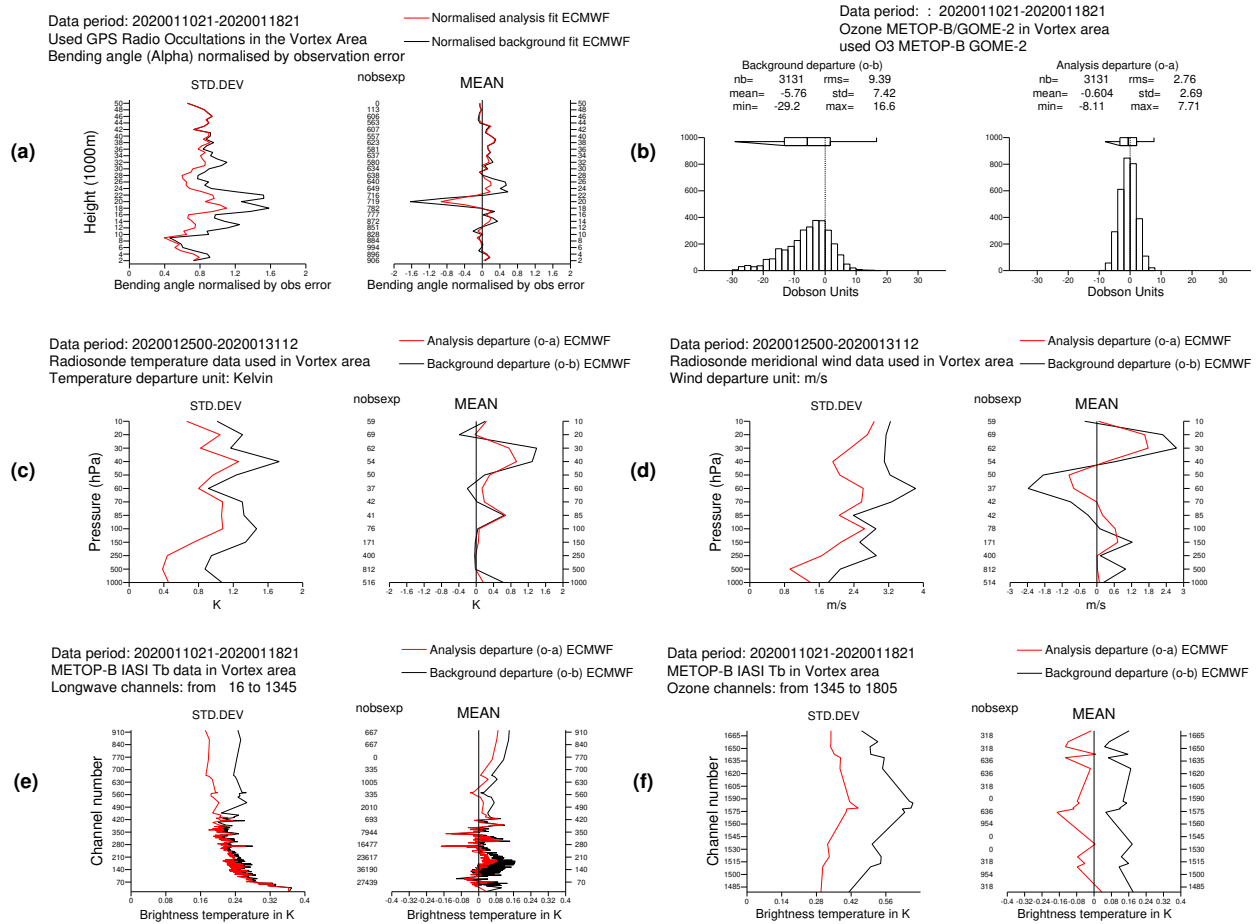

Supplementary Figure 6: **Forcing by assimilation of the observations in the IFS**
(a) Right panel: mean GPS-RO bending angle departure in the area of the vortex normalised by the observation error as a function of altitude. In black, the departure from the background (prior to the assimilation). In red, the departure of the analysis (posterior to the assimilation). The column on the left indicates the number of observations. Left panel: prior and posterior standard deviation. (b) Probability density distribution of the METOP-B/GOME-2 departure in the area of the vortex with respect to the background (left panel) and with respect to the analysis (right panel). (c) Same as (a) but for the radiosonde temperature using pressure as vertical axis. (d) Same as (c) but for the radiosonde meridional wind. (e) Same as (a) but for the METOP-B IASI brightness temperature in the longwave channels. The vertical axis shows the channel number. (f) Same as (e) but for the brightness temperature in the ozone channels. Data averaged over the period 11-18 January for panels (a), (b), (e) and (f). Data averaged over the period 25-31 January for panels (c) and (d).

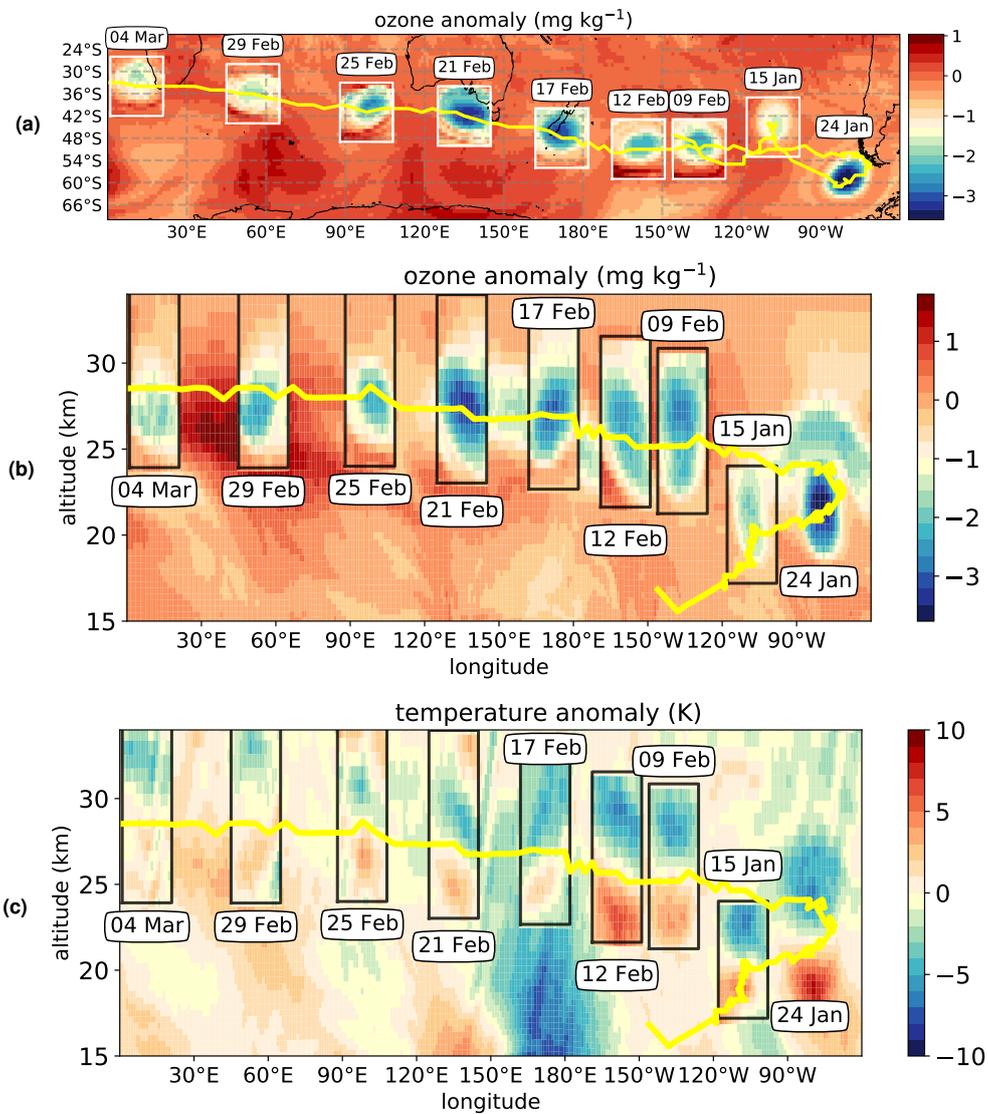

Supplementary Figure 7: **Ozone and temperature**
Upper panel a): horizontal composite chart of the ozone anomaly. Mid panel b): Vertical longitude-altitude composite section of the ozone anomaly. Bottom panel c): Vertical longitude-altitude composite section of the temperature anomaly. These panels are built exactly like Fig. 6a-b of the main text. The anomalies are defined according to the zonal average at the same time. When averaged in time the composite image of Fig. 6c of the main text is obtained.

|         | January | February |
|---------|---------|----------|
| TOA     | $-0.17\pm0.07$ W/m$^2$ | $-0.31\pm0.09$ W/m$^2$ |
| Surface | $-0.66\pm0.13$ W/m$^2$ | $-0.98\pm0.17$ W/m$^2$ |

Supplementary Table 1: **Radiative Forcing**
Area-weighted global-equivalent RF of Australian fire perturbations to the straospheric aerosol layer, at TOA and surface, for January and February 2020.

# Supplementary Note 1    Radiative forcing

The equinox-equivalent daily-average shortwave surface and top of the atmosphere (TOA) direct radiative forcing (RF) are estimated using the UVSPEC (UltraViolet SPECtrum) radiative transfer model in the libRadtran (library for Radiative transfer) implementation [1] and a similar methodology as in [2, 3]. All radiative parameters are computed in the shortwave range 300 to 3000 nm, at 0.1 nm spectral resolution, based on the input solar spectrum of [4]. The background atmospheric state is set using the AFGL (Air Force Geophysics Laboratory) summer mid- or high-latitudes climatological standards [5], depending on the latitude range. As a first-order reference estimate, clear-sky conditions are considered. A shortwave surface albedo of 0.07, typical of sea surface (most of the fire plume disperse over ocean), is used [6]. The RF is very sensitive to the surface albedo and using a fixed albedo value might introduce uncertainties to our estimations. For the radiative forcing calculations, a baseline simulation is first carried out, with the mentioned setup and average aerosols extinction coefficient (750 nm) profiles from OMPS, for January and February 2019. A fire-perturbed run is then performed, using the measured Australian fire aerosols extinction coefficient (750 nm) profiles from OMPS, for January and February 2020. The OMPS USask retrievals are used, for both baseline and fire-perturbed periods. The spectral variability of the aerosol extinction, during fire-perturbed periods, is modelled using the measured Ångström exponent from SAGE III, for January 2020. Typical non-perturbed values of the Ångström exponent are used for the baseline period. Only perturbations of the optical properties of stratospheric aerosol are considered in the present study (OMPS profiles are averaged only using altitude levels from tropopause plus 1 km, in order to avoid possible cloud perturbations). Different perturbed runs have been performed using three different values of the single scattering albedo (0.85, 0.90 and 0.95) and a Heyney-Greenstein phase function with an asymmetry parameter of 0.70. These are typical values of biomass burning aerosol optical properties [e.g. 3, 7, and references therein]. For both the baseline and fire plumes configurations, we run multiple times the radiative transfer simulations at different solar zenith angles (SZA). The daily-average shortwave TOA clear-sky RF for the fire-perturbed aerosol layer is calculated as the difference in SZA-averaged upward diffuse irradiance between the baseline simulation (i.e. without the considered aerosols) and the simulation with aerosols, integrated over the whole shortwave spectral range. The shortwave surface clear-sky RF is calculated as the difference in the SZA-average downward global (direct plus diffuse) irradiance between the simulation with aerosols and the baseline simulation, integrated over the whole spectral range. The clear-sky RF is estimated for both January and February 2020, by comparing with baseline layers of January and February 2019, respectively. Different latitude bands are considered separately, 15 to 25°S, 25 to 60°S and 60 to 80°S. We exclude the latitude band 80 to 90°S because OMPS observations are not available in this band. It is supposed that the Australian fires have no impact in the northern hemisphere.

The equinox-equivalent daily-average clear-sky RF at TOA and surface due to Australian fire perturbations to the stratospheric aerosol layer, averaged over the months of January and February, is shown in Supplementary Figure 2, for the three latitude bands mentioned above. In the latitude band between 25 and 60°S, an average monthly radiative forcing (RF) as large as about -1.0 $W\,m^{-2}$ at TOA and -3.0 $W\,m^{-2}$ at the surface is found in February 2020, that can be attributed to the Australian fires plumes perturbation of the stratospheric aerosol layer. Lower RF values are found at more northern and southern latitude bands, in the Southern Hemisphere. The RF in January are 30 to 50% lower than in February. Based on these clear-sky estimations with detailed radiative transfer calculations, all-sky RF can be derived using the average monthly cloud fraction, in the same latitude bands defined above, and the parameterisation of [8]. Typical cloud fractions in the area affected by the plume are between 40 and 50%, at lower latitudes, up to more than 70%, at the highest latitudes in the Southern Hemisphere [9]. Based on that, an all-sky RF reduced to about 50% of the clear-sky RF, is likely.

Using these regional (latitudinal-limited) RF estimations, we calculate the corresponding area-weighted global-equivalent RF, for January and February 2020. Values are given in Supplementary Table 1, for the detailed clear-sky reference runs. The area-weighted global-equivalent clear-sky RF due to the Australian fires plumes peaks in February, with values as large as -0.31±0.09 $W\,m^{-2}$, at TOA, and -0.98±0.17 $W\,m^{-2}$, at the surface. Values as large as -0.2 $W\,m^{-2}$, at TOA, and -0.5 $W\,m^{-2}$, at the surface, are expected in all-sky conditions. Due to the mentioned spatial limitation of our analyses (we exclude a possible impact in the northern hemisphere and between 80 and 90°S), these global-equivalent clear-sky RF might be slightly underestimated. The all-sky RF, being obtained with a simple parameterisation in terms of the integrated cloud fraction, must be taken with caution and are to be considered a first estimation of the overal RF of Australian fires 2019/2020. Work is ongoing to explicitly simulate the radiative impact of clouds, in our simulations, and to incorporate measured clouds properties.

# Supplementary Note 2   Late evolution of the smoke bubble from CALIOP

On 7 March, the aerosol bubble was over the Atlantic and entered the region of the South Atlantic Anomaly where CALIOP retrieval is too noisy to allow a detection [10]. It emerged about one week later in the Pacific where it could be detected again from 14 March. In the mean time, it was still followed by OMPS (see Fig. 10c of the main text). Supplementary Figure 3 shows a series of CALIOP sections after this date from 16 March until 4 April 2020. Between 28 March and 1 April the bubble split in two parts due to vertical shear, as it occurred by the end of February. The top part was followed for a few more days until 4 April when it reached 55°E after which it was lost in CALIOP data. The bottom part travelled westward but slower and dispersed over the Indian Ocean. By 13 April, a number of patches could still be seen between 30 and 32 km and 25°S-30°S over the Indian Ocean (not shown).

Supplementary Figure 5d shows that the IFS analysed vortex after 16 March is much weaker than during the period displayed in Fig. 5a of the main text. It is also less compact and stands out much less relative to the environment, as shown by the extension of the half maximum vorticity contour in Supplementary Figure 3. By 1 April a collocated vorticity maximum could still be detected in the IFS analysis, but could hardly be considered as a vortex. In other words, even if the presence of a vortical structure is suggested from the persisting confinement, the IFS fails to track it as the temperature disturbance gets too weak to be identified by the observing systems. This is consistent with the loss of the GPS-RO temperature signal seen in Fig. 5d of the main text.

# Supplementary Note 3   Vortices and tracer confinement as seen from TROPOMI

On the top of TROPOMI AI data (see Methods) used to follow the aerosol plume, we also look at CO and $O_3$ tracers to investigate the impact of the fires on stratospheric gaseous composition.

## Supplementary Note 3.1   CO columnar content

TROPOMI provides total CO vertical columns, exploiting clear-sky and cloudy-sky Earth radiance measurements. The retrieval of the CO content is based on the Shortwave Infrared CO Retrieval (SICOR) algorithm [11–13]. The algorithm takes in account the sensitivity of the CO measurements to the atmospheric scattering due to cloud presence employing a two-stream radiative transfer solver. The algorithm is an evolution of the CO retrieval algorithm used for the Scanning Imaging Absorption Spectrometer for Atmospheric Chartography (SCIAMACHY)[14] but with a specific improvement in the CO retrieval for cloudy and aerosol loaded atmospheres. The CO total column densities are retrieved simultaneously with effective cloud parameters (cloud optical thickness and cloud center height) by means of a scattering forward simulation. The inversion exploits the monthly vertical profiles of CO, spatially averaged on a 3x2 degrees grid, from the chemical transport model TM5 [15] for a profile scaling approach. The algorithm has been extensively tested against SCIAMACHY, which covers the same spectral region with the same spectral resolution as TROPOMI but with a lower signal-to-noise ratio, lower radiometric accuracy, and lower spatial resolution [16, 17]. More details on the CO columnar product can be found in [18].

## Supplementary Note 3.2   Ozone columnar content

The TROPOMI Ozone level 2 offline data are retrieved using the GODFIT (GOME-type Direct FITting) algorithm. This is based on the tuning of the simulated radiances in the Huggins bands (fitting window: 325-335 nm) by varying some of the key atmospheric parameters in the state vector to better fit the observations. These parameters include the total ozone, the effective scene albedo and the effective temperature. This approach gives improved retrievals accuracy with respect to the classical Differential Optical Absorption Spectroscopy approach under extreme geophysical conditions as large ozone optical depths. In the offline version, the data are then filtered and kept only if specific criteria are fullfilled (total column density positive but less than 1008.52 DU, respective ozone effective temperature variable greater than 180 K but less than 260 K, ring scale factor positive but less than 0.15, effective albedo is greater than -0.5 but less than 1.5 [19]. More details on the algorithm and on the quality of the datasets can be found in [20–22].

## Supplementary Note 3.3   Aerosol and tracers confinement in the vortex

Supplementary Figure 4 shows the anomaly in the atmospheric composition linked to the vortex on two different days after the plume injection (17 January and 3 February in 4a and 4b, respectively). On both days the satellite images indicate an alteration of the atmospheric composition at the same location as the vortex position identified by the IFS. Supplementary Figure 4a captures the vortex 18 days after the plume injection. Besides the clear aerosol confinement inside the vortex, the compact structure is very visible in the CO total column with values that are enhanced in the vortex with respect to the surrounding background ($2 \cdot 10^{22}$ mol·m$^{-2}$ with respect to a background lower than $1 \cdot 10^{22}$ mol·m$^{-2}$). The IFS vortex centroid is on this day located at 19.8 km and the $O_3$ depletion caused by the atmospheric composition perturbation is also visible on the total column, with a reduction of 60 DU with respect to the surroundings (that reach values of around 330 DU). On the 3 of February (15 days after) the vortex is higher in altitude (centroid from IFS at 24.3 km) and still keeps its confinement power. The vorticty patch still matches the high AI value and CO columnar content enhancement, although with lower gradients with respect to the previous days. This is due to the natural dilution, decaying and leaking of the tracers and the aerosol content. The ozone mini hole, on the other hand, has increasing amplitude with respect to the previous days, with a reduction of 75 DU with respect to the surroundings (around 325 DU). In this day the ozone hole was the most evident, mostly due to the increase on the surrounding $O_3$ values. In the following days the vortex kept gradually losing the aerosol and tracer confinement toward the last days of February when TROPOMI lost the ability to clearly distinguish the vortex in both tracers and aerosol images. This corresponds to a split of the aerosol bubble in two parts as observed by CALIOP (see Fig. 5a of the main text)

### Supplementary Note 3.4  Hemispheric impact of the aerosol plume

Supplementary Figure 4c shows the time composite of the daily TROPOMI AI from 6 January, the day following the second big injection of the smoke plumes. The chart depicts the time evolution of the aerosol plumes released in the atmosphere and clearly shows that the perturbation affected a large fraction of the southern hemisphere. Most of the plume spread and dispersed over the Pacific ocean portion between Australia and South America. Nevertheless parts of the plumes detached from the main one, forming various bubbles of aerosol and sometimes generating vortices that travelled all around the hemisphere (see Fig. 10c of the main text). The daily track of the aerosol centroid for the two main bubbles shows indeed how both moved in different directions and speeds, transporting part of the injected material through all the longitudes. While the main vortex moved slowly, perturbing mostly the southern Pacific for around one month and half, the second vortex made a complete tour around the globe in about three weeks. In addition to those main events, other smaller bubbles were identified and are discussed in the main text (Figs. 8 and 9).

## Supplementary Note 4  Vortex in the IFS

### Supplementary Note 4.1  Time evolution

Supplementary Figure 5 shows the temporal evolution of the vortex in the IFS analysis and of the aerosol bubble from CALIOP data. The results illustrate how the bubble trajectory was closely followed with no significant deviation during the whole period. The path from 4 January to 1 April 2020 completed more than one and a half times the Earth's circumference (66,000 km during 88 days). The time evolution of the potential temperature and of the vorticity are displayed in Fig. 7 of the main text.

### Supplementary Note 4.2  Forcing by assimilation of the observations

Numerical weather prediction requires good knowledge of the initial state of the atmosphere, land and ocean to provide the starting point for the forecast model. This is achieved by the data assimilation process that adjusts a short-range forecast (typically 6-12 hours) to be in closer agreement with available observations. The ECMWF is using 4-dimensional variational analysis method [23] that uses around 25 million observations each 12-hours to perform this adjustment towards the true state of the atmosphere. Supplementary Figure 6 shows examples of the rttude/longitude box around the vortex. Supplementary Figure 6 panels a and c-f show the vertical distribution of observation-minus-background (o-b, black curves) and observation-minus-analysis (o-a, red curves). The left panels show the random component (standard deviation) and the right panels show the systematic error (bias). The data assimilation scheme is extracting useful information from the observations if the random errors are smaller in the analysis (red curves) and/or the bias is reduced (closer to to zero line). It is evident that the analysis of the vortex is improved from using the data from GPS radio-occultations (panel a), radiosonde temperatures and winds from Falkland Islands (panel c,d) and IASI radiances sensitive to long wave temperatures (panel e) and ozone (panel f). Panel b shows that GOME-2 ozone measurements from METOP also contributed to the improved ozone analysis near the vortex.

# Supplementary Note 5  Ozone and temperature

Supplementary Figure 7a-b shows the evolution of ozone in the ECMWF analysis in the same way as Fig.3a-b of the main text. Ozone is depleted with respect to the environment by as much as 3.5 mg kg$^{-1}$ (or 2.1 ppmv, see also Fig. 3c of the main text). We saw above in Supplementary Section Supplementary Note 4.2 that this depletion is maintained by the assimilation of satellite observations. Like vorticity, the ozone distribution sways according to the deformations of the vortex and keeps an ovoid shape in the vertical plane during all the displayed period.

Supplementary Figure 7c shows that the temperature distribution maintains instead a dipolar structure, with a cold pole above and a warm pole below, during the entire period of the vortex evolution. It is noticeable that in the displayed sections it is the separation line rather that the axis between the warm and the cold pole that aligns with the vortex vorticity and ozone axis.